\documentclass[preprint2]{aastex}

\usepackage{color}
\usepackage{float}
\slugcomment{Resubmitted to Astrophysical Journal in May 2015}

\shorttitle{The SU Aur Disk}
\shortauthors{de Leon et al.}

\begin{document}


%\bibliographystyle{astron}

\title{Near-IR High-Resolution Imaging Polarimetry of the SU Aur Disk: Clues for Tidal Tails?\footnote{Based on data collected at Subaru Telescope, which is operated by the National Astronomical Observatory of Japan.}}


\author{
Jerome de Leon\altaffilmark{1,2},
Michihiro Takami\altaffilmark{1},
Jennifer L. Karr\altaffilmark{1},
Jun Hashimoto\altaffilmark{3,4},
Tomoyuki Kudo\altaffilmark{5},
Michael Sitko\altaffilmark{6},
Satoshi Mayama\altaffilmark{2},
Nobuyuki Kusakabe\altaffilmark{4},
Eiji Akiyama\altaffilmark{4},
Hauyu Baobab Liu\altaffilmark{1},
Tomonori Usuda\altaffilmark{4},
Lyu Abe\altaffilmark{7}, 
Wolfgang Brandner\altaffilmark{8}, 
Timothy D. Brandt\altaffilmark{9},
Joseph Carson\altaffilmark{10}, 
Thayne Currie\altaffilmark{5},
Sebastian E. Egner\altaffilmark{5},
Markus Feldt\altaffilmark{8},
Katherine Follette\altaffilmark{11}, %
Carol A. Grady\altaffilmark{12,13},
Miwa Goto\altaffilmark{14},
Olivier Guyon\altaffilmark{5},
Yutaka Hayano\altaffilmark{5},
Masahiko Hayashi\altaffilmark{4},
Saeko Hayashi\altaffilmark{5},
Thomas Henning\altaffilmark{8}, %
Klaus W. Hodapp\altaffilmark{15}, %hawaii
Miki Ishii\altaffilmark{4}, %naoj
Masanori Iye\altaffilmark{4}, %naoj
Markus Janson\altaffilmark{16}, %stockholm
Ryo Kandori\altaffilmark{4}, %naoj
Gillian R. Knapp\altaffilmark{9}, %princeton
Masayuki Kuzuhara\altaffilmark{17}  %4,25,26}, %Titech
Jungmi Kwon\altaffilmark{18}, %Todai dept Astro
Taro Matsuo\altaffilmark{19}, %Kyoto
Michael W. McElwain\altaffilmark{13}, %
Shoken Miyama\altaffilmark{20}, %Hiroshima
Jun-Ichi Morino\altaffilmark{4}, %naoj
Amaya Moro-Martin\altaffilmark{21},  %CAB
Tetsuo Nishimura\altaffilmark{5}, %subaru
Tae-Soo Pyo\altaffilmark{5}, %subaru
Eugene Serabyn\altaffilmark{22},  %JPL
Takuya Suenaga\altaffilmark{2}, %soken
Hiroshi Suto\altaffilmark{4}, %naoj
Ryuji Suzuki\altaffilmark{4},  %naoj
Yasuhiro Takahashi\altaffilmark{18,23}, %todai dep. astro, mext
Naruhisa Takato\altaffilmark{5},  %subaru
Hiroshi Terada\altaffilmark{5}, %subaru
Christian Thalmann\altaffilmark{24}, %amsterdam
Daigo Tomono\altaffilmark{5}, %subaru
Edwin L. Turner\altaffilmark{9,25}, %princeton, kavli
Makoto Watanabe\altaffilmark{26}, %hokkaido
John P. Wisniewski\altaffilmark{3}, %oklahoma
Toru Yamada\altaffilmark{27}, %tohoku
Hideki Takami\altaffilmark{5}, %subaru
Motohide Tamura\altaffilmark{4,5,18} %todai, naoj, subaru
}
\altaffiltext{1}{Institute of Astronomy and Astrophysics, Academia Sinica, P.O. Box 23-141, Taipei 10617, Taiwan, R.O.C.; jpdeleon.bsap@gmail.com} % Jerome, Hiro, Jennifer, Baobab
\altaffiltext{2}{The Center for the Promotion of Integrated Sciences, The Graduate University for Advanced Studies~(SOKENDAI), Shonan International Village, Hayama-cho, Miura-gun, Kanagawa 240-0193, Japan} % Jerome, Mayama, Suenaga, Tamura
\altaffiltext{3}{H.L. Dodge Department of Physics and Astronomy, University of Oklahoma, 440 W Brooks St Norman, OK 73019, USA} %Hashimoto, Wisniewski
\altaffiltext{4}{National Astronomical Observatory of Japan, 2-21-1 Osawa, Mitaka, Tokyo 181-8588, Japan} % Eiji, Hashimoto, Kusakabe, Currie, Tamura, etc
\altaffiltext{5}{Subaru Telescope, 650 North A'ohoku Place, Hilo, HI 96720, USA} % Kudo, Egner
\altaffiltext{6}{Department of Physics, University of Cincinnati, Cincinnati OH 45221, USA} % Sitko
\altaffiltext{7}{Laboratoire Lagrange (UMR 7293), Universit\`{e} de Nice-Sophia Antipolis, CNRS, Observatoire de la C\^{o}te d'Azur, 28 avenue Valrose, 06108 Nice Cedex 2, France} % Abe
\altaffiltext{8}{Max Planck Institute for Astronomy, Koenigstuhl 17, D-69117 Heidelberg, Germany} %Brandner, Feldt
\altaffiltext{9}{Department of Astrophysical Sciences, Princeton University, Peyton Hall, Ivy Lane, Princeton, NJ 08544, USA}
%Brandt
\altaffiltext{10}{Department of Physics and Astronomy, College of Charleston, 58 Coming St., Charleston, SC 29424, USA} % Carson
\altaffiltext{11}{Kavli Institute of Particle Astrophysics and Cosmology, Stanford University, 452 Lomita Mall, Stanford, CA 94305, USA.} % Follette
\altaffiltext{12}{Eureka Scientific, 2452 Delmer Suite 100, Oakland CA 96402, USA} % Grady
\altaffiltext{13}{ExoPlanets and Stellar Astrophysics Laboratory, Code 667, Goddard Space Flight Center, Greenbelt, MD 20771, USA} % Grady, McElwain
\altaffiltext{14}{Universit\"ats-Sternwarte M\"unchen, Scheinerstr. 1, D-81679 Munich, Germany} % Goto
\altaffiltext{15}{Institute for Astronomy, University of Hawaii, 640 North A'ohoku Place, Hilo, HI 96720, USA} % Klaus
\altaffiltext{16}{Department of Astronomy, Stockholm University, 106 91, Stockholm, Sweden} % Janson
\altaffiltext{17}{Department of Earth and Planetary Sciences, Tokyo Institute of Technology, 2-12-1 Ookayama, Meguro-ku, Tokyo 152-8551, Japan} %Kuzuhara, Toukoudai
\altaffiltext{18}{Department of Astronomy, The University of Tokyo, 7-3-1 Hongo, Bunkyo-ku, Tokyo 113-0033, Japan} % Takahashi, Tamura
\altaffiltext{19}{Department of Astronomy, Kyoto University, Kitashirakawa-Oiwake-cho, Sakyo-ku, Kyoto, Kyoto 606-8502, Japan} % Matsuo
\altaffiltext{20}{ Hiroshima University, 1-3-2 Kagamiyama, Higashi-Hiroshima, 739-8511, Japan}
\altaffiltext{21}{Department of Astrophysics, CAB-CSIC/INTA, 28850 Torrejon de Ardoz, Madrid, Spain} % Moro-Martin
\altaffiltext{22}{Jet Propulsion Laboratory, California Institute of Technology, Pasadena, CA, 91109, USA} % Serabyn
\altaffiltext{23}{MEXT, 3-2-2- Kasumigaseki, Chiyoda, Tokyo 100-8959} %Takahashi
%\altaffiltext{24}{Astronomical Institute "Anton Pannekoek", University of Amsterdam, Postbus 94249, 1090 GE, Amsterdam, The Netherlands} %Thalmann
\altaffiltext{24}{Institute for Astronomy, ETH Zurich, Wolfgang-Pauli-Strasse 27, 8093, Zurich, Switzerland} % Thalmann
\altaffiltext{25}{Kavli Institute for the Physics and Mathematics of the Universe, The University of Tokyo, Kashiwa 277-8568, Japan} % Turner
\altaffiltext{26}{Department of Cosmosciences, Hokkaido University, Kita-ku, Sapporo, Hokkaido 060-0810, Japan} % Watanabe
\altaffiltext{27}{Astronomical Institute, Tohoku University, Aoba-ku, Sendai, Miyagi 980-8578, Japan} % Yamada
%\altaffiltext{28}{Department of Earth and Planetary Science, The University of Tokyo, 7-3-1 Hongo, Bunkyo-ku, Tokyo 113-0033, Japan} %Kuzuhara, Todai
%\altaffiltext{29}{Institute for Astronomy, ETH Zurich, Wolfgang-Pauli-Strasse 27, 8093, Zurich, Switzerland} % Thalmann before?

%%%%%%%%%%%%%%%%%%%%%%%%%%%%%%%%%%%%%%
%%%%%%%%%%%%%%%%%%%%%%%%%%%%%%%%%%%%%%
%                                      Section 0 (Abstract)
%%%%%%%%%%%%%%%%%%%%%%%%%%%%%%%%%%%%%%
%%%%%%%%%%%%%%%%%%%%%%%%%%%%%%%%%%%%%%

\begin{abstract}
We present new high-resolution ($\sim$0\farcs09) $H$-band imaging observations of the circumstellar disk around the T Tauri star SU Aur. Our observations with Subaru-HiCIAO have revealed the presence of scattered light as close as 0\farcs15 ($\sim$20 AU) to the star. Within our image, we identify bright emission associated with a disk with a minimum radius of $\sim$90 AU, an inclination of $\sim$35$\degr$ from the plane of the sky, and an approximate P.A. of 15$\degr$ for the major axis. We find a brightness asymmetry between the northern and southern sides of the disk due to a non-axisymmetric disk structure. We also identify a pair of asymmetric tail structures extending east and west from the disk. The western tail extends at least 2\farcs5 (350 AU) from the star, and is probably associated with a reflection nebula previously observed at optical and near-IR wavelengths. The eastern tail extends at least 1\arcsec (140 AU) at the present signal-to-noise.
%We estimate that the dust mass of the tail is significantly lower than that of the disk inferred from millimeter observations.
%We discuss the possibility that an encounter with an unseen brown dwarf %or a free-floating planet 
%has produced such tails.
These tails are likely due to an encounter with an unseen brown dwarf, but our results do not exclude the explanation that these tails are outflow cavities or jets.
\end{abstract}



%%%%%%%%%%%%%%%%%%%%%%%%%%%%%%%%%%%%%%
%%%%%%%%%%%%%%%%%%%%%%%%%%%%%%%%%%%%%%
%                                    Section 1 (Introduction)
%%%%%%%%%%%%%%%%%%%%%%%%%%%%%%%%%%%%%%
%%%%%%%%%%%%%%%%%%%%%%%%%%%%%%%%%%%%%%

\section{INTRODUCTION}
%For a long time, planets are believed to form in circumstellar disks around pre-main sequence stars.
Direct images of scattered light from protoplanetary disks provide valuable information about their surface geometries.
%In recent years, the number of such observations for pre-main sequence (PMS) stars has been steadily increasing.
Advances in instrumentation have produced high angular resolution, high sensitivity observations, revealing a variety of substructures in protoplanetary disks associated with pre-main sequence (PMS) stars.  In particular, near-infrared (IR) imaging polarimetry with adaptive optics has been used extensively during the Strategic Explorations of Exoplanets and Disks with Subaru (SEEDS) survey with %Subaru-
HiCIAO \citep[][]{Tamura09}. This survey has been successful in observing diverse disk morphologies, from uniform \citep[e.g., MWC 480---][]{Kusakabe12} to complex, including spirals \citep[e.g., AB Aur, SAO 206262, MWC 758---][]{Hashimoto11,Muto12,Grady13}, gaps or rings \citep[e.g., PDS 70, 2MASS J1604, Oph IRS 48---][]{Hashimoto12,Mayama12, Follette15}, and other asymmetric flux distributions \citep[e.g., UX Tau A, SR 21, RY Tau, Oph IRS 48---][]{Tanii12,Follette13,Takami13,Follette15}. Some of these structures may be due to dynamical disk-planet interactions providing hints of ongoing planet formation
\citep[e.g.,][]{Hashimoto11,Hashimoto12,Muto12,Mayama12,Grady13}.
%Moreover, there has been considerable progress in theoretical modelling of spiral structures in disks \citep[see][for a review]{Papaloizou_PPV} which, in the context of spiral density waves, are also explained to be due to unseen planet(s) within the disk \citep{Hashimoto11,Muto12,Grady13}.

%Second, state the need for your work, as an opposition between what the scientific community currently has and what it wants. //Since the first optical polarimetric observations by Elsasser and Staude (1978), there exists only a handful of directly imaged circumstellar disks around PMS stars. 
%Adding to the catalogue of detailed near-IR coronagraphic imaging observations conducted for low-mass pre-main sequence stars (T Tauri stars) with interesting disk morphology,
SU Aur is a classical T Tauri star located in the Taurus-Auriga star formation region at a distance of $\sim$140 pc \citep{Bertout06}.
This star-disk system shows large IR and UV excesses indicating the existence of a circumstellar disk and ongoing active mass accretion, respectively
\citep[e.g.,][]{Bertout88,Calvet04,Jeffers14}.
%The star also shows strong photometric variability caused by dark starspots and bright accretion hotspots (e.g., DeWarf et al. 2003).
Table \ref{table:suaur_params} summarizes its stellar and disk properties. 

SU Aur is also associated with a bright reflection nebula \citep{Nakajima95,Grady01,Chakraborty04,Jeffers14}. \citet{Nakajima95} revealed an associated cometary nebulosity at optical wavelengths, extending to the west and southwest at small ($<$10\arcsec) and large distances (10\arcsec-30\arcsec), respectively. Optical high-resolution observations using coronagraphy or imaging polarimetry \citep[]{Grady01,Chakraborty04,Jeffers14} have resolved the extended nebulosity toward the west within $5\arcsec$ of the star. Possible origins of the observed structure include an outflow cavity \citep[][]{Grady01,Chakraborty04}, part of the circmustellar disk \citep[][]{Chakraborty04},
or part of a remnant envelope or parent cloud \citep[]{Jeffers14}. \citet{Chakraborty04} also applied point-spread function (PSF) subtraction to seeing-limited near-IR images, and report the presence of an extended nebulosity to the southwest, close to the star.

%In this paper, we shed light on the nature and origin of the extended nebulosity associated with the disk of SU Aur by investigating the morphology of structures found in the inner regions ($< 1$\arcsec) for the first time in the near-IR. 
In this paper, we present near-IR imaging polarimetry of %around the star
SU Aur. Using Subaru-HiCIAO, we have revealed scattered light from the disk, as well as two tails which may be due to tidal interaction with an unseen brown dwarf. %or a free-floating planet. 
The rest of the paper is organized as follows.
In section 2, we summarize our observations and data reduction. In section 3, we show the observed Polarized Intensity (PI) flux distribution and polarization. We highlight the discovery of salient features %and structures 
including the disk and tail structures. In section 4, we discuss the orientation of the disk and the nature of the tail structure. 

%%%%%%%%%%%%%%%%%%%%%%%%%%%%%%%%%%%%%%
%%%%%%%%%%%%%%%%%%%%%%%%%%%%%%%%%%%%%%
%                         Section 2 (Observations and Data Reduction)
%%%%%%%%%%%%%%%%%%%%%%%%%%%%%%%%%%%%%%
%%%%%%%%%%%%%%%%%%%%%%%%%%%%%%%%%%%%%%

\section{OBSERVATIONS AND DATA REDUCTION}

SU Aur was observed in $H$-band (1.65 \micron) on 2014 January 19 UT using the high-contrast imager %imaging instrument 
HiCIAO.
% \citep{Tamura06_HiCIAO,Hodapp08,Suzuki10} as part of the SEEDS survey. %mounted on the Subaru Telescope. 
%The observations were conducted as part of the SEEDS survey. 
The adaptive optics system (AO188)
%\citep[AO188,][]{Hayano04,Minowa10} 
provided a stable stellar PSF (FWHM = 0$\farcs$08, Strehl ratio = 0.2--0.3). We used a combination of polarimetric and angular differential imaging (PDI+ADI) modes. A dual Wollaston prism was used to split incident light into two %pairs of images 
image pairs that are orthogonally and linearly polarized, each with a 5$\arcsec$$\times$5$\arcsec$ field of view and a pixel scale of 9.5 mas pixel$^{-1}$. As for several other SEEDS observations, linear polarization %(i.e., Stokes parameters) 
was measured by rotating the half-wave plate to four angular positions: 0$\degr$, 22.5$\degr$, 45$\degr$, and 67.5$\degr$. We obtained 12 full waveplate cycles, taking a 30-sec exposure per waveplate position, using a 10$\%$ neutral density (ND) filter. No coronagraphic mask was used in order to image the innermost region around the central star. Four additional sets of waveplate cycles were obtained using a 1$\%$ ND filter to measure the PSF FWHM and the total Stokes $I$ flux density of the star ($\it{I_*}$). %The angle of total field rotation was small ($\sim$8$\degr$) during the observations.  The total integration time of the $PI$ image was 1440-sec considering all quality images with small FWHMs (FWHM $<$ 0$\farcs$1) by careful inspections of the stellar PSF.

The data were reduced using the standard method for ADI+PDI \citep{Hinkley09}.
%as used for SEEDS studies \citet[][]{Hashimoto12,Muto12,Takami13}. 
%The reduction was done using self-written codes in pyRAF and and python. %\footnote[1]{a python package based on Image Reduction and Analysis Facility (IRAF)} 
The measured Stokes  $\it{I}$, $\it{Q}$, and $\it{U}$ parameters were used to compute the more accurate intrinsic Stokes parameters.
%$\it{I}$, $\it{Q}$, and $\it{U}$.
We then calculated the polarized intensity (PI $= \sqrt{(Q^2 + U^2)}$), polarization (PI/$I$) and the angle of polarization ($\theta_{P}=$ 0.5$\times$arctan($U$/$Q$)). The instrumental polarization of HiCIAO at the Nasmyth platform was corrected %with/ by
following %the Mueller matrices technique
\citet{Joos08} with errors of $<$0.1$\%$. 

To increase the signal-to-noise, we convolved the $Q$ and $U$ images with a Gaussian having a FWHM of 0\farcs04.
As a result, the angular resolution of the images shown in later sections is 0\farcs09, slightly larger than the instrumental resolution.

We also applied the LOCI algorithm \citep{Lafreniere07} to the Stokes $I$ images to search for faint companions, with parameters as in Table \ref{table:LOCI}.
The image rotation during the observations was small ($\sim$8$\degr$) hence we needed a small $N_\sigma$ (the minimum displacement distance in the azimuthal direction) to search for a companion close to the star (down to $r$=0\farcs17 and 0\farcs36 for $N_\sigma$=0.1 and 0.5 $\times$ FWHM of the observations, respectively). The other LOCI parameters were the same for the four sets and were standard and optimum for SEEDS observations.
We embedded artificial point sources at different radii and derived upper limits for 3-$\sigma$ fluxes of the companion of $m_H$=$\sim$15, $\sim$19, 21.0 and 21.5 at $r$=0\farcs16, 0\farcs5, 1\arcsec, and 2\arcsec~from the star, respectively. The detection limits were almost identical between different parameter sets.

%We also applied the LOCI algorithm \citep{Lafreniere07} to the Stokes $I$ images to search for faint companions. The parameters for subtracting the flux from the parent stars were the size of the optimization area ($N_A$), the width of the subtraction area ($dr$), the ratio of radial and azimuthal width for the optimization area ($g$), and the minimum displacement distance in the azimuthal direction ($N_\sigma$). We tried four parameter sets with different $N_\sigma$ shown in Table \ref{table:LOCI}. The image rotation during the observations was small ($\sim$8$\degr$) hence we needed a small $N_\sigma$ to search for a companion close to the star (down to $r$=0\farcs17 and 0\farcs36 for $N_\sigma$=0.1 and 0.5 $\times$ FWHM of the observations, respectively). The other LOCI parameters were the same between the four sets and were standard and optimum for SEEDS observations. We found that the detection limits were almost identical between different parameter sets and derived upper limits for 3-$\sigma$ fluxes of the companion of $m_H$=14.9, 18.9, 21.0 and 21.5 at $r$=0\farcs16, 0\farcs5, 1\arcsec, and 2\arcsec~from the star, respectively. Since we did not detect a companion, we will focus our discussion on extended structures seen in the $PI$ image for the rest of the paper.

%%%%%%%%%%%%%%%%%%%%%%%%%%%%%%%%%%%%%%
%%%%%%%%%%%%%%%%%%%%%%%%%%%%%%%%%%%%%%
%                         Section 3 (Results)
%%%%%%%%%%%%%%%%%%%%%%%%%%%%%%%%%%%%%%
%%%%%%%%%%%%%%%%%%%%%%%%%%%%%%%%%%%%%%

\section{RESULTS}
%(1) Describe an overview of the bright emission (i.e., the component associated with the disk) 
Figure~\ref{fig:PI} shows the observed PI flux distribution overlaid with polarization vectors. There is bright emission at the center with a radius of $\sim$100 AU. This emission probably originated from a protoplanetary disk elongated 
%at the apparent PAs $\sim$15$\degr$ and $\sim$105$\degr$ northeast. 
to the north-south, slightly offset in the counter-clockwise direction.
%The peak emission located at $r\sim$0\farcs20 ($\sim$28 AU) north of the star has a surface brightness (SB) in $PI$ of 2.9$\times$10$^{-6}$ pix$^{-1}$ ($\sim$80 mJy arcsec$^{-2}$). 
The peak emission in the disk is located at $r$$\sim$0\farcs20 ($\sim$28 AU) north of the star and has PI/$I_*$ of 2.9$\times$10$^{-6}$ pix$^{-1}$. This translates to surface brightness in PI of $\sim$80 mJy arcsec$^{-2}$.
%\textbf{The peak emission in the disk is located at $r\sim$0\farcs20 ($\sim$28 AU) north of the star and has a surface brightness (SB) in $PI$ of $\sim$80 mJy arcsec$^{-2}$ ($PI$/$I_*$=2.9$\times$10$^{-6}$ pix$^{-1}$).}
%This is brighter than the region in the southern side of the disk by a factor of $\sim$2. 
This is comparable to the brightest disks in the SEEDS observations summarized in \citet{Takami14}.
Moreover, the distribution shows a deficit of emission perpendicular to the semi-major axis similar to PI distributions observed in other disks with intermediate inclination angles \citep[e.g.,][]{Hashimoto12,Kusakabe12,Tanii12}.
This deficit likely arises because along the minor axis, the light is dominated by forward- and backward-scattered photons, which are intrinsically less polarized \citep[e.g.,][]{Takami13}, and does not require an actual deficit of material in the disk.
We detect a brightness asymmetry between the disk's northern and southern sides: the northern side is brighter than the southern side by a factor of 2. 
%This is because of either a non-axisymmetric disk structure or illumination \citep[e.g.,][]{Wisniewski08,Hashimoto12,Takami13,Takami14}.
%where the eastern side of the disk is inclined to Earth.

% (2) Then describe the elliptical fitting at the next paragraph. Provide disk orientation, inclination and a lower limit of the disk radius.
To analyze the inclination and orientation of SU Aur's disk, %we fit an isophotal flux to the scaled $PI$ image. The outermost contours, defined by a minimum value ($PI$/${\it I_*}$ = 10$^{-7}$) are used to fit an ellipse to the $PI$ image by a simple visual analysis. 
we fit an ellipse to the outermost contour (PI/$I_*$ = 10$^{-7}$ pix$^{-1}$) by a simple visual analysis. This fit yields angular separations of 1\farcs7 and 1\farcs4 ($\sim$240 AU and $\sim$200 AU) for the major and minor axes, respectively, with a position angle (P.A.) of $\sim$15$\degr$ for the major axis.
%The lower limit of the radial extent of the disk can be traced from $\sim$0\farcs15 to 0\farcs80 which translates to $\sim$21 - 115 AU (projected separation without taking into account inclination) from the star {\bf at} a distance of 140 pc. The geometric center of the fitted ellipse was offset by $\sim$4.2 AU which is consistent with SU Aur's position within errors. 
The ratio between the minor and major axes indicates an approximate disk inclination of $35\degr$.
%, consistent with the disk model for the spectral energy distribution by \citet{Jeffers14}.
We do not make a more quantitative or accurate measurement for the disk orientation or inclination because of the asymmetric geometry seen in the contour.
%Note that an objective ellipse-fitting method in this case would be inappropriate due to SU Aur's complex disk geometry.

\begin{figure}
\centering
\includegraphics[width=7.5cm]{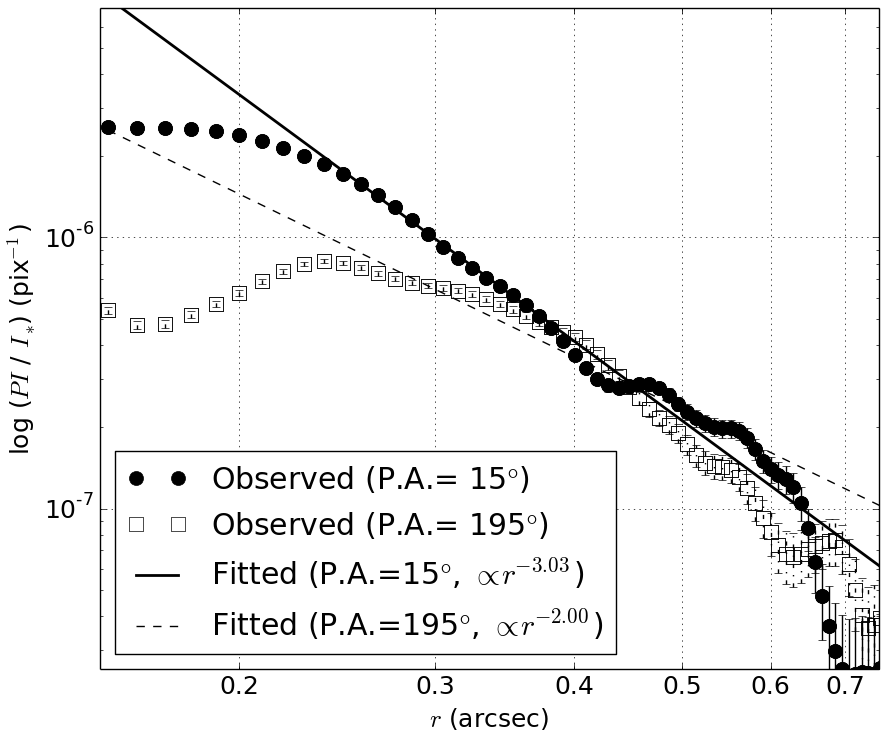}
\caption{Radial PI/$I_*$ profiles along the semi-major axis of the disk.
The filled circles and solid squares show the measured value at P.A.s of 15$\degr$ and 195$\degr$, respectively. The solid and dashed lines show power-law fits made at $r$=0\farcs25--0\farcs6, where $r$ is the projected distance to the star.
\label{fig:radialPI_strip}}
\end{figure}

% (3) Radial profile fitting
We extracted the radial PI/$I_*$ profiles wherein we %and averaged the surface brightness across 
averaged a swath of 11 pixels ($\sim$0\farcs1) along the disk's major axis %from the boundary of the software mask until the disk's edge
on both sides of the star.
A power-law function was fitted to each profile, as shown in Figure~\ref{fig:radialPI_strip}.
Between 0\farcs25 and 0\farcs6, we find the following power-law indices that best fit the observed PI profiles: $\sim$--3 and $\sim$--2 at P.A. of $15\degr$ and $195\degr$, respectively.
The %power-law 
indices we measured are roughly consistent with SEEDS studies of several disks \citep[SAO 206462, MWC 480, MWC 758 ---][]{Muto12,Kusakabe12,Grady13} as well as some other observations of disks associated with Herbig Ae and Be stars \citep{Fukagawa10}.
%\textbf{The observed profile at P.A. of $15\degr$ follows the fit rather well, implying a generally smooth surface distribution. The observed profile at P.A. of $195\degr$, however, shows apparent deviations wherein the slope appears to be steeper ($<$ --3) close to the disk edge while it appears flatter ($>$--2) at radii close to the star, implying a relatively more complex geometry. The discrepancy in indices supports the idea of nonaxisymmetric illumination or disk structure between the northern and southern sides of the disk.}
Compared with P.A. = $15\degr$, the observed profile shows a larger deviation from the power-law fit at P.A.=195\degr (i.e., shallower and steeper slopes at inner and outer radii, respectively). A different brightness distribution between these two P.A.s indicates that the disk structure is non-axisymmetric.

%Note that the azimuthal profile of the disk was not reported because of the deficit in the $PI$ flux. %Taking into account the f(r)$\sim$r$^{-2}$ drop in stellar radiation, this imples the surface density of the scattering dust grains drops off by about $\Sigma$ $\propto$ r$^{-1}$ if we assume constant scattering properties of the dust grains over the observed radial range (Quanz et al. 2011).
% This slope, roughly, is observed in several other HAeBe disks (e.g., Fukagawa et al. 2010; Quanz et al. 2011). // SU Aur's index is steeper except PDS 70 and 2MASS J1604 with indices -3.2 and -4 (at least). 

Figure~\ref{fig:radialPI_strip} also shows a drop-off in the observed flux at 0\farcs6--0\farcs7 from the star. This indicates a lower limit for the disk radius of $\sim$90 AU at a distance of 140 pc. This is consistent with disk models for millimeter emission and optical-to-millimeter spectral energy distributions by \citet{Ricci10} (100-300 AU) and \citet{Jeffers14} (500 AU), respectively. Note that the absence of flux in the outer region does not imply the absence of disk material as it may be due to self-shadowing of the disk \citep{Takami14}.

% (4)overview of faint extension from the disk, including the spirals and tails. After the overview, describe their details.

%\subsubsection{Tail Associated with the Disk}
In Figure~\ref{fig:PI}, a long, faint 
tail (PI/$I_*$ $\lesssim$ 0.5$\times 10^{-7}$ or 1.4 mJy arcsec$^{-2}$)
extending westward at least 2\farcs5 (350 AU) can also be seen. 
We also find an extended component that is marginally detected from the north-east to the east side of the disk.
The tail in the west is likely associated with 
%optical imagery as an extended ``cometary'' nebulosity \citep{Nakajima95} or extended circumstellar environment \citep{Jeffers14}, and in the near-IR coronagraphic imaging as radially streaming filamentary structure \citep{Chakraborty04}. 
the extended reflection nebula previously observed at optical and near-IR wavelengths \citep{Nakajima95,Grady01,Chakraborty04,Jeffers14}.
However, the distribution of emission appears significantly narrower in the north-south direction than previous observations in the optical and near-IR.
%two structures are quite different since near-IR observations by \citet{Chakraborty04} probed a different region in the disk than that observed by \citet{Jeffers14} based on the P.A. of strongest emission. %The tail in our image lies at P.A. $\sim250\degr$
%corresponding to regions of little emission in the optical image \citep[e.g.,][]{Jeffers14, Nakajima95}. %Moreover, \citet{Chakraborty04} measured the extent of the tail to be 3\farcs5 in $J$ band and 2\farcs5 in the K band between P.A.$=210\degr$ and $270\degr$ consistent with our observations.
%Within 1$\arcsec$ of the star, our observations clearly show the tail's inner extension which appears to be associated with a tidal tail in the south of the disk as explained as follows.

To better display structures in the faint outer regions, we convolve the PI image with a Gaussian of FWHM=0\farcs1 and then scale the PI flux at each pixel by $R^2$, where $R$ is the projected distance from the star (Figure~\ref{fig:tail}).
%We select the \textbf{power-law index $\alpha$ of 2} to best clarify the extended emission from the disk.
The figure clearly shows two tails
%spiral-like arms, one wrapping 
extending east and west.
%These tail structures are asymmetric with %S1 
%the western tail is wider and brighter by a factor of $\sim$2 than %S2
%the eastern tail.
In Figure~\ref{fig:tail}, the tails appear to be associated with the edge of the disk emission, however, observations with higher signal-to-noise are required to confirm this.

Figure~\ref{fig:PI} also shows the polarization vector map. 
%The vectors show a centrosymmetric pattern, as has been observed in several other disks with near-IR imaging polarimetry \citep[e.g.,][]{Hashimoto11}. This indicates that the disk and tail emission are both scattered light that originated from the central star.
The vectors in the disk show a centrosymmetric
pattern, as has been observed in several other
disks with near-IR imaging polarimetry \citep[e.g.,][]{Hashimoto11}. This indicates that the
disk is seen in scattered light that originated from the central star. Similarly, the vectors in the tail indicate that it is illuminated by the star-disk system.
The disk has lower fractional polarization ($\lesssim$10$\%$) than the western tail (up to $\sim$30$\%$). This is because the Stokes $\it{I}$ flux within 1$\arcsec$ of the star is dominated by the star's PSF halo, lowering the degree of polarization (PI/$I$). Hence, the vectors represent the lower limit of the true degree of polarization. 

%%%%%%%%%%%%%%%%%%%%%%%%%%%%%%%%%%%%%%
%%%%%%%%%%%%%%%%%%%%%%%%%%%%%%%%%%%%%%
%                         Section 4 (Discussion)
%%%%%%%%%%%%%%%%%%%%%%%%%%%%%%%%%%%%%%
%%%%%%%%%%%%%%%%%%%%%%%%%%%%%%%%%%%%%%
\section{DISCUSSION}

\begin{figure*}[h!]
\centering
%\vspace{-20cm}
\includegraphics[width=17cm]{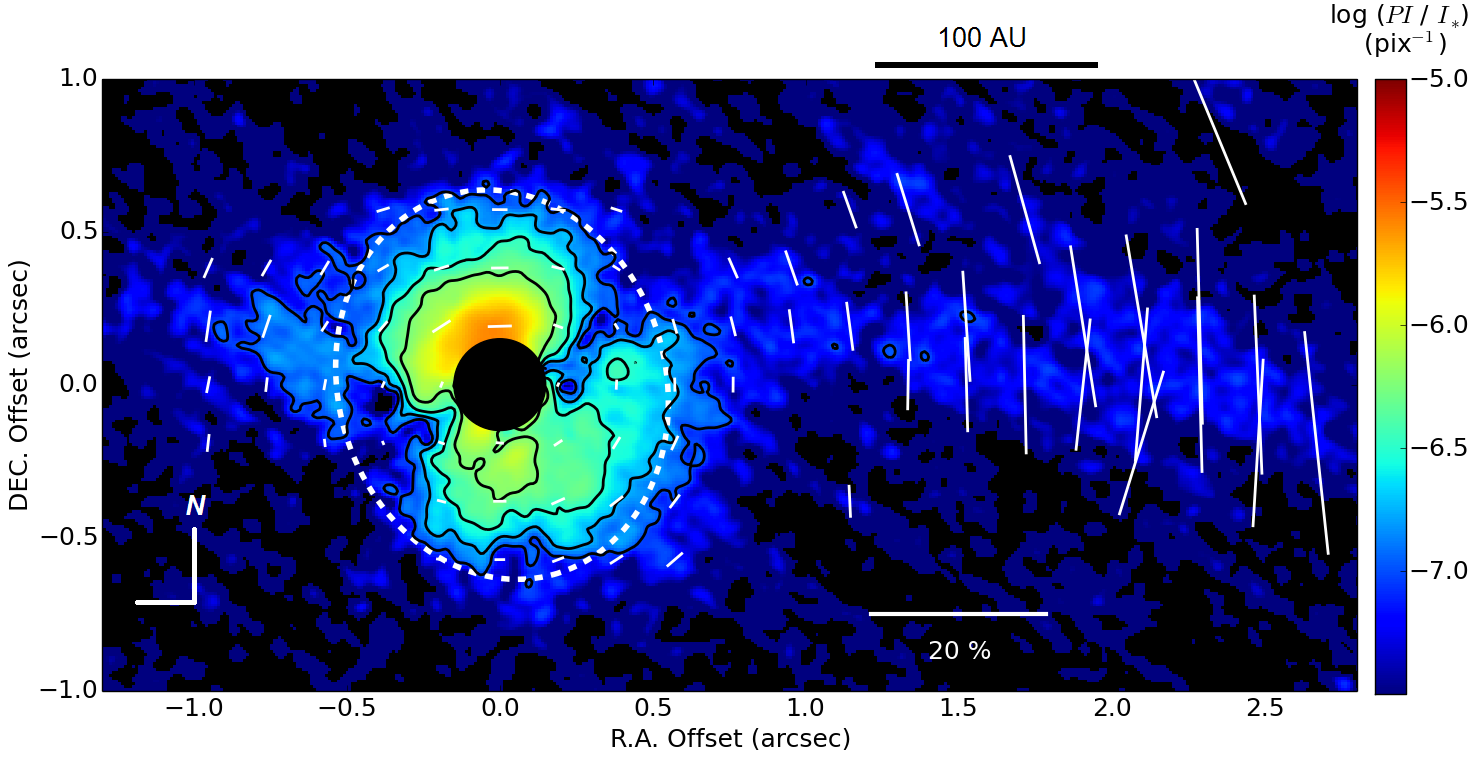}
\caption{PI image of SU Aur in log-scale overlaid with polarization vector map. 
The PI flux at each pixel is scaled by the spatially integrated stellar $I$ flux (PI/$I_*$$=$10$^{-7}$ corresponds to 2.8 mJy arcsec$^{-2}$).
%\textbf{To measure the disk angular size and inclination of SU Aur's disk, the outermost contour is used to fit an ellipse (white dashed lines) by visual analysis (see text for details)}.
%The white dot shows the ellipse center offset. 
The contours are shown for a log scale of PI/$I_*$ of --7.0, --6.75, --6.5 and --6.25.  The dotted white ellipse is adjusted to the outermost contour (see text for details).
The region within $r=$0\farcs15 ($\sim$20 AU) is masked to show only regions where the measurements are reliable. %Triangles denote the opening of the flux deficit due to depolarization along the semi-minor axis. 
%The lengths and directions of vectors indicate the degree and angle of polarization, respectively. 
\label{fig:PI}}
\end{figure*}

\citet{Jeffers14} conducted optical imaging polarimetry of SU Aur and found a disk-like morphology elongated in the east-west direction at a larger angular scale (radius of $\sim$2$\arcsec$). The disk-like structure within 1\arcsec of the star seen in the near-IR (Figure \ref{fig:PI}) is almost perpendicular to this structure in the optical. %The discrepancy between the two wavelengths may be due to different contributions from a remnant envelope surrounding the star-disk system. 
%As the dust opacity is higher in the optical than in the near-IR, scattered light associated with a remnant envelope can contribute more at the optical wavelengths. 
As the dust opacity is higher in the optical than in the near-IR, the disk may be partially obscured by dust in the envelope at optical wavelengths. Thus, near-IR observations should have an advantage in observing scattered light close to the disk surface \citep{Fukagawa04}. This explanation is supported by the fact that the power-law indices of the radial PI distribution in the near-IR are similar to other disk systems (Section 3).

%\citet{Akeson05} has measured the orientation of the major axis of about 115\degr ~at miliarcsec scales using near-IR interferometry. This is, again, nearly perpendicular to the disk orientation shown in Figure \ref{fig:PI}. Different measurements by \citet{Akeson05} may be due to non-axisymmetric structures due to planet formation. A better signal-to-noise and $u-v$ coverages may be required to confirm their measurements.

\begin{figure}
\centering
\includegraphics[width=7.5cm]{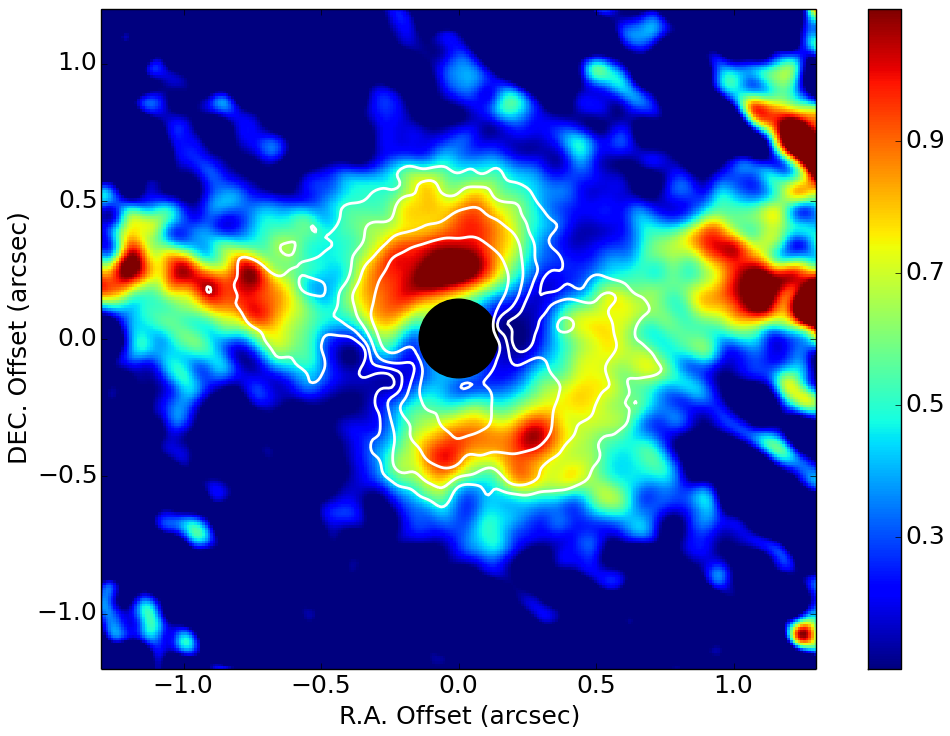}
\caption{
The outer disk and tails, highlighted by scaling the PI flux at each pixel by $R^{2}$, where $R$ is the projected distance to the star.
The image is shown within 1\farcs3 of the star, where this image processing does not significantly enhance background fluctuation.
The color scale is arbitrary. The white contours show the original PI flux distribution shown in Figure \ref{fig:PI}.
 \label{fig:tail}}
\end{figure}

Tails like those in Figures \ref{fig:PI} and \ref{fig:tail} are not usually observed in scattered light associated with young stellar objects (YSOs). An exception may be a jet-like feature in the Z CMa system observed by \citet{Millan-Gabet02}, which the authors attributed to a cavity wall. In the remaining section we discuss possible origins of these tails:%, i.e., 
(1) outflow cavities, (2) collimated jets, and (3) tidal tails due to a (sub-)stellar encounter.

Before discussing the nature of the SU Aur tails in detail, we estimate the dust mass of the western tail. \citet{Takami13} derived the following equation to estimate the dust mass observed in scattered light using imaging polarimetry:
\begin{equation}
m_{dust}=\int \frac{n_{PI}({\bf r})}{n_*} \frac{r^2}{\kappa_{\rm ext}} \Big(\frac{PI}{I_0} \Big)^{-1}~d\bf{r}
\end{equation}
where $m_{dust}$ is the dust mass; $n_{PI} (\bf r)$ is the number of PI photons at each position; $n_*$ is the number of stellar $I$ photons; $r$ is the distance to the star; $\kappa_{ext}$ is the dust opacity; PI/$I_0$ is the fraction of the PI flux normalized to the incident flux on the dust grains. Here we assume the interstellar dust size distribution measured by \citet{Kim94}. This yields $\kappa_{ext}=5.3 \times10^3$  cm$^2$ g$^{-1}$ and PI/$I_0 \sim 0.01$ at a wavelength of 1.65 \micron ~\citep{Takami13}. Integrating Equation (1) from disk edge to the tail end in Figure \ref{fig:PI} ($\sim$110 -- 350 AU) yields a dust mass of $\sim$6 $\times10^{-8} ~M_{\Sun}$, 
% =(5.972E24 *0.02)/1.989E30
%$\sim$ 2 $\times 10^{-2} ~M_{\Earth}$}
if we use the projected distance for $r$ at the distance of 140 pc. This mass is smaller than the disk's dust mass %of the 
inferred from millimeter interferometry (see Table \ref{table:suaur_params}) by a factor of 2000--5000. %1000--3000
However, like protoplanetary disks, a significantly larger mass might exist behind the scattering layer. The above dust mass should therefore be a lower limit for the tail.

SU Aur is known to be associated with a reflection nebula in the east-west direction, approximately the same direction as the tails in Figures \ref{fig:PI} and \ref{fig:tail}, and to the southwest at larger angular scales \citep{Nakajima95,Grady01,Chakraborty04,Jeffers14}. Such nebulosity associated with YSOs is often attributed to an outflow cavity \citep[e.g.,][]{Tamura91,Hodapp94,Lucas98,Padgett99}.
However, this explanation may not apply to these tails for several reasons.
First, the tails shown in Figures \ref{fig:PI} and \ref{fig:tail} are significantly narrower than other YSO outflow cavities, which generally have a wide opening angle. Secondly, an outflow cavity in scattered light is usually observed toward a single direction unless we see the disk edge-on \citep[e.g.,][]{Fischer94,Whitney03b}. However, the tails in Figures \ref{fig:PI} and \ref{fig:tail} are seen in two opposite directions despite the fact that the disk has an intermediate viewing angle.

Another possible explanation for the tails may be that these are collimated jets. Many YSOs are associated with such jets \citep[e.g.,][for a review]{Frank14},  however, these are not usually observed in scattered light. To further investigate this explanation, we estimate the mass ejection rate and compare it with the disk accretion rate. Using Equation (1) we derive a dust mass of 3$\times$10$^{23}$ g AU$^{-1}$ along the western tail at 200 AU from the star. This would yield a mass ejection rate of 2$\times$10$^{-7}~M_{\Sun}$ yr$^{-1}$, assuming a typical jet velocity of 50 km s$^{-1}$ \citep[e.g., ][]{Hartigan95,Hirth97} and gas-to-dust mass ratio of 100. This is 7--40 times larger than the disk accretion rate measured by \citet{Calvet04} and \citet{Ricci10} (see Table \ref{table:suaur_params}). The mass ejection to mass accretion ratio would be even larger if the tail contains more gas and dust than is inferred from the scattered light. However, such a high mass ejection to mass accretion ratio is not feasible for PMS stars, which usually show a mass ejection to mass accretion ratio of $\sim$0.1 \citep[e.g.,][]{Calvet97}. Therefore, jets are not likely to explain the tail structure either, unless the jet velocity is significantly lower (e.g., $\lesssim$10 km s$^{-1}$) than for other similar objects.

Tidal interaction with a star or a brown dwarf may alternatively explain the observed tail structures. Kinematic simulations of stellar encounters show that this physical process produces tails similar to those in Figures \ref{fig:PI} and \ref{fig:tail}  \citep[e.g,][]{Pfalzner03,Forgan09}. A tidal tail was observed in the RW Aur disk system using millimeter interferometry \citep{Cabrit06}. A difficulty with this interpretation is that there is yet no observational evidence of another star within 30$\arcsec$ of SU Aur \citep[][see also data archive for 2MASS]{Nakajima95,Grady01,Chakraborty04,Jeffers14}. However, an encounter with an unseen brown dwarf (BD) %or a free-floating planet 
could produce tails %whose mass is significantly larger than the disk but
observable in scattered light.
%an upper limit to the brightness of a possible companion based on their I image. Can you rule out a late M-star for example?
%
%\textbf{
%Using an evolutionary model for sub-stellar objects (i.e., COND model; Baraffe et al. 2003), we estimate the mass of the putative brown dwarf to be 6-7$\times$10$^{-4}$ $M_{\odot}$ for 1 and 10 Myr, respectively, or approximately 2$\times$10$^{-3}$ $M_{\odot}$ corresponding to our observations' detection limit of 21.0$-$21.5 mag %for r=1\arcsec and r=1\farcs5, respectively
%(see section 2).
%The same models and the completeness limit of 2-Micron All Sky Survey Point Source Catalog (2MASS PSC; 15.1 mag) yield the upper limit masses of 0.8$\times$10$^{-2}$ $M_{\odot}$ and 2$\times$10$^{-2}$ $M_{\odot}$, respectively.}
We estimate an upper limit BD mass of 0.6-2$\times 10^{-3}$ $M_{\odot}$ for our field of view of observations with a detection limit of 21.0-21.5 mag (Section 2) and the COND models for 1-10 Myr \citep{Baraffe03}. The same models and the completeness limit of the 2MASS Point Source Catalog (15.1 mag) yield an upper limit mass of 0.8-2$\times$10$^{-2}$ $M_{\odot}$.

A (sub-)stellar encounter would affect the disk structure, and therefore affect mass accretion and planet formation \citep[e.g,][]{Pfalzner03,Forgan09,Forgan10,Breslau14}.
%This physical process should occur more frequently in young stellar clusters, which are responsible for formation of a significant fraction of stars in our Galaxy, therefore it may affect disk accretion and planet formation for more systems.
SU Aur would be a good, rare observational example to test these theories in detail. Confirmation of this physical process in this system would require deep imaging to search for a nearby brown dwarf, %or a free-floating planet as well as 
and kinematic information for the disk and the tail from millimeter interferometry.
A search for jet emission at optical and near-IR wavelengths would also be a good test for the jet hypothesis \citep{Frank14}.

%%%%%%%%%%%%%%%%%%%%%%%%%%%%%%%%%%%%%%
%%%%%%%%%%%%%%%%%%%%%%%%%%%%%%%%%%%%%%
%                         Section 5 (Conclusions)
%%%%%%%%%%%%%%%%%%%%%%%%%%%%%%%%%%%%%%
%%%%%%%%%%%%%%%%%%%%%%%%%%%%%%%%%%%%%%
\section{CONCLUSIONS}

%The Conclusion section presents the outcome of the work by interpreting the findings at a higher level of abstraction than the Discussion and by relating these findings to the motivation stated in the Introduction.
We present $H$-band imaging polarimetry of scattered light from the circumstellar disk around the T Tauri star SU Aur. Our high-resolution (0\farcs09) image of polarized intensity has allowed us to observe scattered light from the disk as close as 0\farcs15 ($\sim$20 AU) to the star. We identify bright emission probably associated with the disk, elongated along an approximate P.A. of 15$\degr$. From its distribution, we estimate a minimum disk radius of $\sim$90 AU and a disk inclination of $\sim$35$\degr$.
%We find significant brightness asymmetry between the northern and southern sides of the disk along the major axis. The surface brightness distribution %$S(r)$ in the northern side of the disk follows $\sim$$r^{-3}$, as observed in several protoplanetary disks. In the southern side of the disk, however, %$S(r) \propto \sim$$r^{-2}$ the distribution follows $\sim$$r^{-2}$ with significant deviations between 0\farcs2 and 0\farcs65, where $r$ is the projected distance to the star.
As observed in several protoplanetary disks, the surface brightness along the disk's major axis is approximately proportional to $r^{-2}$ to $r^{-3}$ at $r$=0\farcs2--0\farcs65, where $r$ is the projected distance to the star.
The brightness distribution of the scattered light is asymmetric between the disk's northern and southern sides indicating a non-axisymmetric disk structure.

Our near-IR images also show two tails extending to the east and west of the disk. 
The western tail extends at least 2\farcs5 (350 AU) from the star, and is probably associated with an extended reflection nebula previously observed at optical and near-IR wavelengths.
%However, the width of the extension in the north-south direction is significantly narrower than the other observations.
%The eastern tail is significantly shorter (1\arcsec--1\farcs5) than the western tail at a given signal-to-noise.
The eastern tail extends at least 1\arcsec (140 AU) at the present signal-to-noise.
We estimate a lower limit of the dust mass of the western tail ($r$=110--350 AU) of $\sim$$6 \times 10^{-8} ~M_{\Sun}$ %$\sim$$3 \times 10^{-2} ~M_\Earth$ 
using an interstellar dust model and assuming that the tail lies close to the plane of the sky. This mass is significantly lower than the disk's dust mass inferred from millimeter interferometry ($1-3 \times 10^{-4} M_{\Sun}$).

Possible explanations for the tails include (1) outflow cavities, (2) collimated jets, and (3) tidal interaction with a brown dwarf.
%\textbf{whose upper limit mass is approximately between 2$\times$10$^{-3}$ $M_{\odot}$ and 2$\times$10$^{-2}$ $M_{\odot}$ using the COND model given our observations' detection limit and 2MASS PSC completeness limit, respectively.} %or a free-floating planet. 
%The morphology and brightness suggest that the last explanation is the most likely. 
The morphology and brightness favor the last explanation, but these do not rule out the other possibilities.
Confirmation requires searching for a nearby brown dwarf with deep imaging, and observing the kinematics of the disk and tails using millimeter interferometry.

\acknowledgments
%We are grateful for an anonymous referee for a thorough review and valuable comments.
We thank the Subaru Telescope staff for their support, especially Michael Lemmen for making our observations successful.
We also thank Drs. Kazushi Sakamoto, Shigehisa Takakuwa, Lihwai Lin, Yoichi Ohyama, Pin-Gao Gu and Henry Hsieh for useful discussions.
M.T. is supported from Ministry of Science and Technology (MoST) of Taiwan (Grant No. 103-2112-M-001-029).
C.A.G. acknowledges support under NSF AST 1008440.
%
%%%%%%%%%%%%%%%%%%%%%%%%%%%%%%%%%%%%%%%%%%%%%%%%%%%%%%%%%%%%
%%%
%%%  %%% Comment for Jerome %%%
%%%
%%%  The remaining are copied from the Takami+14 paper. We will edit it after we forward the draft
%%%  to the entire SEEDS team.
%%%
%
%T.M. is supported by JSPS KAKENHI Grant Numbers 26800106, 23103004, 26400224.
%
%R.D. acknowledges the support for this work by NASA through Hubble
%Fellowship grant HST-HF-51320.01-A awarded by the Space Telescope Science
%Institute, which is operated by the Association of Universities for
%Research in Astronomy, Inc., for NASA, under contract NAS 5-26555.
%
%Jungmi Kwon is supported by the JSPS Research Fellowships for Young Scientists (PD: 24$\cdot$110).
%
%J.C. was supported by NSF-AST 1009203. 
%
%
%J.P.W. is supported by NSF-AST 1009314.
%
This research made use of the Simbad database operated at CDS, Strasbourg, France, and the NASA's Astrophysics Data System Abstract Service.

{\it Facilities:} \facility{Subaru (HiCIAO)}.

%\bibdata{/Users/hiro/Desktop/astro.bib}

%\bibliography{/Users/hiro/Desktop/astro}

\clearpage

%% the file, uncomment the \plotone, \plottwo, and
%% If you need a layout that cannot be achieved with \plotone or
%% \plottwo, you can invoke the graphicx package directly with the
%% \includegraphics command or use \plotfiddle. For more information,
%% please see the tutorial on "Using Electronic Art with AASTeX" in the
%% documentation section at the AASTeX Web site,
%% http://www.journals.uchicago.edu/AAS/AASTeX.
%%
%% The examples below also include sample markup for submission of
%% supplemental electronic materials. As always, be sure to check
%% the instructions to authors for the journal you are submitting to
%% for specific submissions guidelines as they vary from

%% This example uses \plotone to include an EPS file scaled to
%% 80% of its natural size with \epsscale. Its caption
%% has been written to indicate that additional figure parts will be
%% available in the electronic journal.

%%%%%%%%%%%%%%%%%%%
%%%%%%%%%%%%%%%%%%%
%  Figure 1 : PI images
%%%%%%%%%%%%%%%%%%%
%%%%%%%%%%%%%%%%%%%

%%%
%%%   %%% Comment for Jerome %%%
%%%
%%%   \begin{figure*} and \end{figure*} are used for figures with a full width of the page.
%%%   \begin{figure} and \end{figure} are used for figures with a half width of the page.
%%%

%
%%%
%

%%%%%%%%%%%%%%%%%%%
%%%%%%%%%%%%%%%%%%%
%  Figure 2 : radial distribution
%%%%%%%%%%%%%%%%%%%
%%%%%%%%%%%%%%%%%%%
\clearpage
%
%%%
%

%%%%%%%%%%%%%%%%%%%%%%%%%
%%%%%%%%%%%%%%%%%%%%%%%%%
%  Figure 3 : Processed figure to better show tails
%%%%%%%%%%%%%%%%%%%%%%%%%
%%%%%%%%%%%%%%%%%%%%%%%%%

%
%%%
%

%% If you are not including electonic art with your submission, you may
%% mark up your captions using the \figcaption command. See the
%% No more than seven \figcaption commands are allowed per page,
%% so if you have more than seven captions, insert a \clearpage

%% Tables should be submitted one per page, so put a \clearpage before

%% Two options are available to the author for producing tables:  the
%% deluxetable environment provided by the AASTeX package or the LaTeX
%% table environment.  Use of deluxetable is preferred.
%%

%% Three table samples follow, two marked up in the deluxetable environment,
%% one marked up as a LaTeX table.

%% In this first example, note that the \tabletypesize{}
%% command has been used to reduce the font size of the table.
%% We also use the \rotate command to rotate the table to
%% landscape orientation since it is very wide even at the
%% Note also that the \label command needs to be placed

%% This table also includes a table comment indicating that the full
%% version will be available in machine-readable format in the electronic

\clearpage

%%%%%%%%%%%%%%%%%%%%%%%%
%%%%%%%%%%%%%%%%%%%%%%%%
%%% Table 1. Parameters for SU Aur
%%%%%%%%%%%%%%%%%%%%%%%%
%%%%%%%%%%%%%%%%%%%%%%%%
\begin{table}
\caption{Parameters for SU Aur \label{table:suaur_params}}
\begin{tabular}{lcc}
\tableline\tableline
Parameter & & Reference \\ \tableline
Spectral Type 		&       G2III        	                                                &	1 \\
Stellar Mass		&	$1.9 \pm 0.1 ~M_{\odot}$                                       & 2 \\
Age				&	$6.8 \pm 0.1$ Myr                                                   & 2 \\
Dust mass of the disk    &	1-3$\times 10^{-4}~M_{\odot}$                        &	3 \\
Mass accretion rate	&	0.5-3$\times 10^{-8} ~M_{\odot}$ yr$^{-1}$ &	4,5 \\
Distance			&	$143^{+17}_{-13}$ pc                                           & 6 \\
$A_V$			&	 0.9 mag                                                             & 2 \\
$H$ mag.			&	6.56 mag                                                                 &    7 \\
\tableline
\end{tabular} \\
\tablerefs{[1] \cite{Herbig52} ; [2] \citet{Bertout07} ; [3] \citet{Ricci10}; [4] \citet{Calvet04}; [5] \citet{Ricci10}; [6] \citet{Bertout06}; [7] 2MASS All Sky Catalog of Point Sources}
\end{table}

%%%%%%%%%%%%%%%%%%%%%%%%
%%%%%%%%%%%%%%%%%%%%%%%%
%%% Table 2. LOCI
%%%%%%%%%%%%%%%%%%%%%%%%
%%%%%%%%%%%%%%%%%%%%%%%%
\begin{table}
\caption{LOCI Parameters\label{table:LOCI}}
\begin{tabular}{lc}
\tableline\tableline
Parameter\tablenotemark{a} & Values\tablenotemark{b} \\ \tableline
$N_A$ & 150 $\times$ FWHM\\
$dr$ & 5 $\times$ FWHM ($r < 50 ~\times$ FWHM) \\
& 100$\times$FWHM ($r > 50 ~\times$ FWHM) \\
$g$ & 1 \\
$N_\sigma$  & 0.1/0.3/0.5/1.0 $\times$ FWHM\\
\tableline
\end{tabular} \\
\tablenotetext{a}{$N_A$ ... the size of the optimization area; $dr$ ... the width of the subtraction area; $g$ ... the ratio of radial and azimuthal width for the optimization area; $N_\sigma$ ... the minimum displacement distance in the azimuthal direction \citep{Lafreniere07}.}
\tablenotetext{b}{FWMH is that of the observations (0\farcs08).}
\end{table}

\end{document}